\documentclass[a4paper,10pt]{article}

\usepackage[utf8]{inputenc}
\usepackage[T1]{fontenc}
\usepackage{lmodern}
\usepackage{graphicx}
\usepackage{amsmath}

\title{Ticket-based multi-strand method for increased efficiency in
  proof-of-work based blockchains}

\author{Elias Rudberg\\
mail@eliasrudberg.se
}

\begin{document}

  \maketitle

  \begin{abstract}
    This paper outlines a method aiming to increase the efficiency of
    proof-of-work based blockchains using a ticket-based approach. To
    avoid the limitation of serially adding one block at a time to a
    blockchain, multiple semi-independent chains are used such that
    several valid blocks can be added in parallel, when they are added
    to separate chains. Blocks are added to different chains, the
    chain index being determined by a ``ticket'' that the miner must
    produce before creating a new block. This allows increasing the
    transaction rate by several orders of magnitude while the system
    is still fully decentralized and permissionless, and maintaining
    security in the sense that a successful attack would require the
    attacker to control a significant portion of the whole network.
  \end{abstract}

  \section{Introduction}

  A proof-of-work blockchain system as described in \cite{Bitcoin2008}
  provides several interesting features. In particular, such systems
  can be:

  \begin{itemize}
  \item Decentralized: no central authority
  \item Permissionless: anyone can decide to participate
  \end{itemize}

  Information is stored in a distributed ledger, copies of which can
  exist in many places but no instance has authority over the
  others. The proof-of-work mechanism provides a form of collective
  decision-making regarding how information can be added to the
  distributed ledger.

  One drawback of such blockchain systems is that the rate at which
  information can be added to the ledger is rather limited. This is
  because information is always added one block at a time, serially,
  at an even pace. For example, for the Bitcoin blockchain a new block
  is added about every 10 minutes, while for Monero the corresponding
  time is 2 minutes~\cite{ZeroToMonero2020}.

  For cryptocurrency applications, each block contains one or more
  transactions, and one way to increase the rate of transactions is to
  increase the maximum allowed block size. However, the block size
  cannot be increased infinitely; the work involved in adding a block
  needs to be feasible. In practice, even if the possibilities of
  increasing block size and reduced block time are taken into account,
  it appears unlikely that blockchains such as Bitcoin and Monero as
  currently designed could handle the transaction rates that would be
  necessary if such cryptocurrencies were to be used by a significant
  portion of the world's population. That would require billions of
  transactions per day, several orders of magnitude more than what
  seems possible due to the current limitation of only adding one
  block at a time, serially.

  In this paper we discuss a way to significantly increase the
  efficiency of proof-of-work blockchains by exploiting parallelism,
  while keeping the system decentralized and permissionless. The
  described method builds on similar ideas as discussed
  in~\cite{ParallelChains2018} but realized in a different way. The
  improvement in throughput compared to a single-strand blockchain
  depends on the chosen amount of parallelism. As an example, a
  multi-stranded chain of 1024 strands would in principle allow 1024
  times as many transactions per day.

  % This paper is organized as follows: in Section~\ref{sec:}

  \section{Method}

  In this section we first describe the standard (single-strand)
  proof-of-work blockchain as proposed in \cite{Bitcoin2008} for
  comparison, and then we describe our proposed multi-strand method.

  \subsection{Standard (single-strand) blockchain}

  In a proof-of-work blockchain as described in \cite{Bitcoin2008}
  each block contains the hash of the previous block, a set of
  transactions, and a nonce:

  \vspace{5mm}
  
  \includegraphics[width=0.99\textwidth]{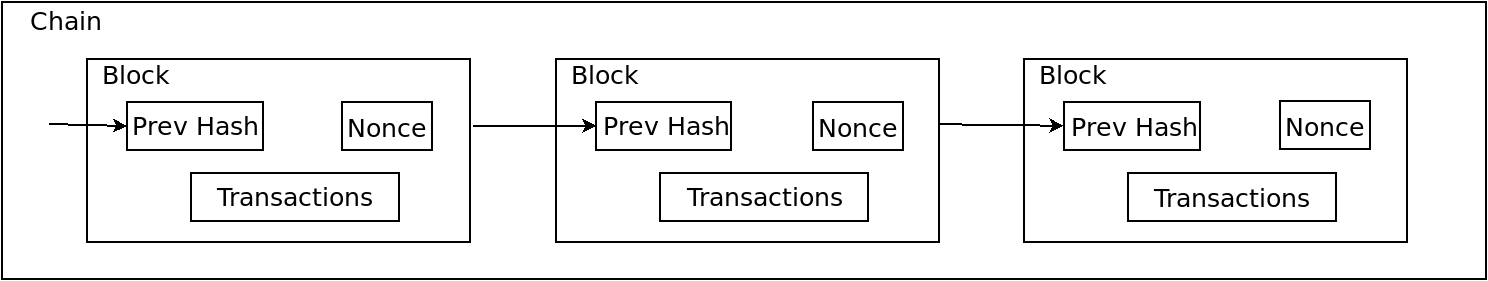}

  \vspace{5mm}
  
  The work of producing a new valid block involves finding a nonce
  such that the hash of the new block begins with a certain number of
  zero bits.

  When a new valid block has been found, that new block is sent to
  other nodes, and the other nodes verify that it is indeed a valid
  block by checking the transactions and the previous hash as well as
  the hash of the new block itself. When other nodes have verified
  that the new block is valid, each of them add it to their local
  chain. They have incentive to do so because the longest chain is
  considered correct, so everyone wants their new blocks to be added
  to a chain that is as long as possible.

  \subsection{Ticket-based multi-strand blockchain} \label{section:ticket_based_method}

  When there is only one chain where a new block can be added, only
  one of the nodes (miners) that are trying to produce a new block can
  succeed, while all the other attempts to create new blocks will
  fail. One obvious way to allow more than one to succeed is to have
  several independent chains; if each chain can grow independently of
  the others, then several valid blocks, one per chain, can be
  produced in parallel. However, having completely independent chains
  would be bad for security since an attacker could pick one chain to
  attack and would only need to redo the proof-of-work for the
  selected chain in order to rewrite history for that chain.

  We want to keep the property that an attacker needs to control a
  large part of the CPU power of the entire network, while we at the
  same allow parallel growth as if we had several independent
  chains. To achieve this, we devise a ``ticket-based'' approach as
  follows.

  Instead of a single chain, we consider a different form of
  distributed ledger that consists of $n = 2^p$ chains which we label
  ``Chain 0'' to ``Chain $n-1$''. Each chain works similarly to the
  single-strand case, with the difference that each block is required
  to include a valid \emph{ticket} and a corresponding signature:

  \vspace{5mm}
  
  \includegraphics[width=0.99\textwidth]{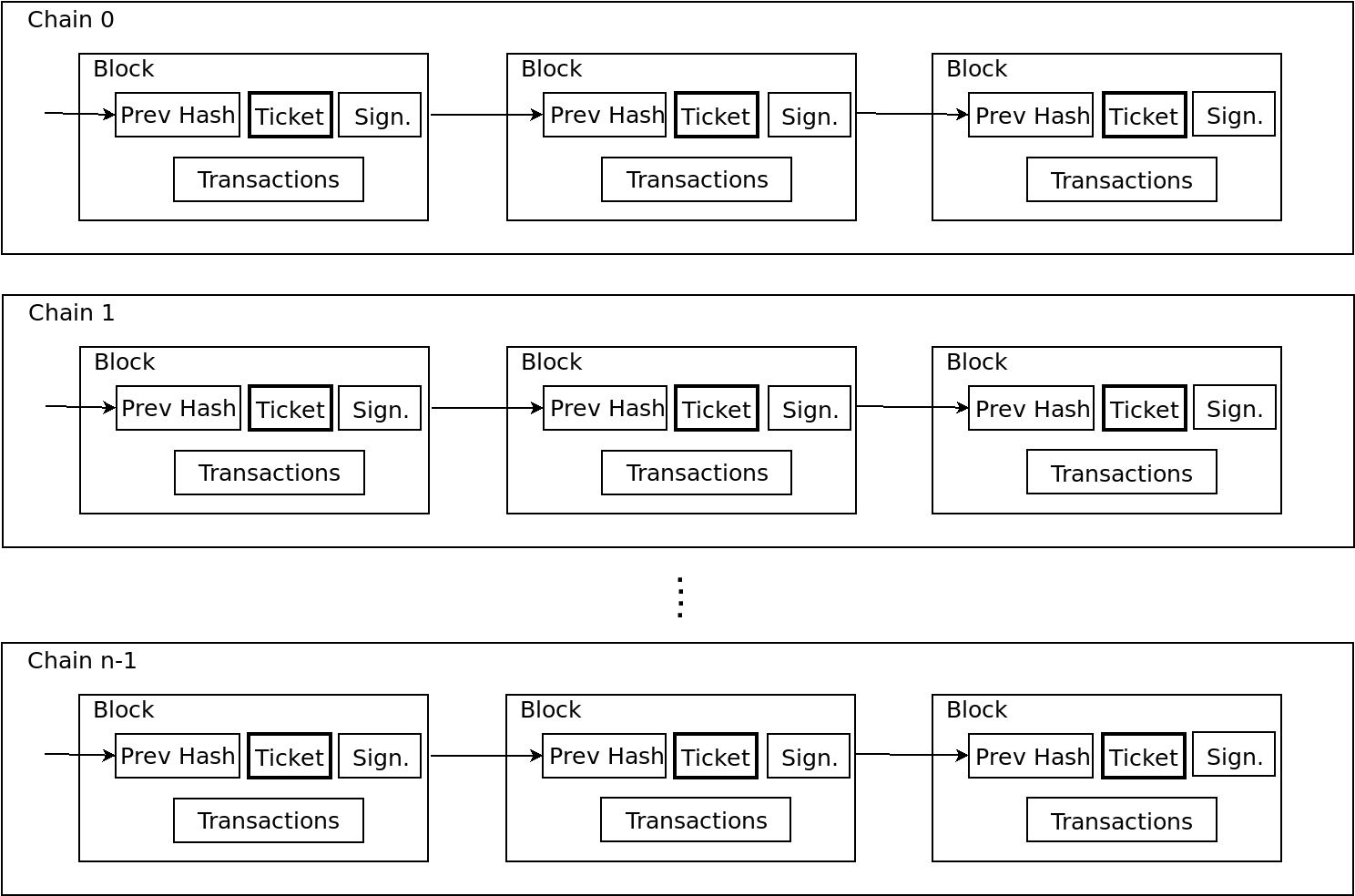}

  \vspace{5mm}
  
  The \emph{ticket} is a data structure that contains the hash of the
  last known block from each of the $n$ chains, together with a pubkey and a nonce
  chosen such that the hash of the ticket satisfies the following: (1)
  the hash starts with a certain number of zero bits and (2) the last
  $p$ bits of the hash correspond to the chain index of the chain for
  which the ticket is valid.

  \vspace{5mm}

  \includegraphics[width=0.77\textwidth]{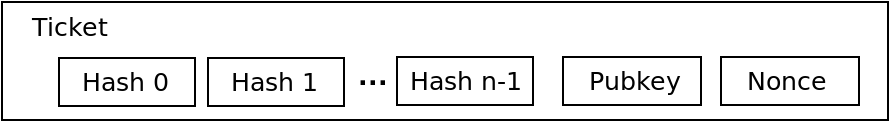}

  \vspace{5mm}

  The signature in each block is used to provide proof that the block
  was indeed created by the ticket-holder. Before creating a ticket,
  the miner will first generate a keypair (private key and
  corresponding pubkey) for the ticket and place the pubkey inside the
  ticket. Later, when a nonce has been found making the ticket valid,
  the miner creates the block and signs it using the private key
  before sending the new block to other nodes. This ensures that
  another node cannot ``hijack'' the ticket and use it to create a
  different block; only the node that created the ticket has the
  ability to create the corresponding block. Note that there should be
  a new keypair for each new ticket and the purpose is to allow others
  to verify the connection between the ticket and the block -- the
  signature shows that whoever created the ticket has also created the
  corresponding block.
  
  The procedure to follow for a miner who wants to add a new block now
  becomes as follows:

  \begin{itemize}
  \item Create a keypair to use for the new ticket
  \item Create a ticket by scanning for a ticket nonce that
    makes the ticket satisfy the criteria (1) and (2) above. Now the
    miner knows which chain it is possible to add a block to.
  \item Then include that ticket in a new block for the chain that the
    ticket is valid for, and add a signature proving that the ticket
    creator has created the block.
  \end{itemize}

  The proof-of-work difficulty is determined by the number of zero
  bits required. The choice of difficulty can be made in the same way
  as for the single-strand case.

  Note that the miner cannot choose which chain to add a block
  to, since the chain index is determined by the ticket. The miner
  tries to find a nonce leading to a hash starting with enough zero
  bits, while some other bits of the same hash determine the chain
  index. The chain index becomes effectively random and the ticket thus
  functions as a lottery ticket giving the miner access to mine a
  block for one specific chain. If a miner were to insist on only
  wanting to add blocks to one specific chain, then that miner would
  need to discard all tickets produced for other chains and the
  probability of getting a ticket with the desired chain index is only
  $1/n$ so the cost in wasted CPU power would be very high. Thus,
  miners have a strong incentive to let the bits of the ticket
  determine the chain to work on, leading to effectively random
  distribution among the $n$ chains. An attacker who wants to target a
  specific chain would still need to control a large part of the CPU
  power of the whole network.

  The procedure to check if a new block is valid is the same as for
  the single-strand case except that the ticket is also verified, as
  follows:

  \begin{itemize}
  \item Calculate the hash of the ticket and check the last $p$ bits to
    find the chain index.
  \item Look at the information stored in the ticket about the hash of
    the last block for that chain index, checking that it in fact
    matches the hash of the last block for that chain.
  \item Check that the required number of zero bits are found in the
    beginning of the hash of the ticket.
  \item Check that the signature for the block is valid, using the
    pubkey given inside the ticket.
  \end{itemize}

  The purpose of requiring the ticket to contain the hash of the last
  block of each chain is to ensure that the ticket is fresh; without
  that requirement it would be possible for an attacker to spend a
  long time hoarding tickets to carry out an attack against a specific
  chain later. The requirement to include the hash of the last block
  of each chain in the ticket makes such hoarding impossible since old
  tickets will no longer be valid.

  Note that when checking if a ticket is valid, only one of the block
  hashes inside the ticket is actually checked, namely the one
  corresponding to the chain index the ticket is for. All the other
  hashes inside the ticket are ignored. The reason we need to include
  all $n$ hashes in the ticket is that when the ticket is created it
  is not yet known which chain index the ticket will be valid for. So
  we include all the $n$ hashes in the ticket, only one of them will
  be the relevant one in the end but it is not known which one until
  later.

  Each of the $n$ chains grows independently of the other chains;
  while the average time to add a block should be the same, there is
  nothing restricting the chains to stay at precisely the same
  length. For example, it may happen that Chain 0 is several blocks
  longer than Chain 1, or the other way around.
  
  To illustrate how the multi-strand method gives increased efficiency,
  consider the following example: we take the simplest case $n=2$ so
  there are two chains, Chain 0 and Chain 1. Imagine there are 100
  different miners. Each miner starts by creating a ticket, and they
  look at the last bit of the ticket hash to determine which chain
  they can mine on. About 50 miners will get Chain 0 and the other 50
  will get Chain 1. One miner succeeds in finding a new block for
  Chain 0, and at about the same time another miner succeeds in
  finding a new block for Chain 1. Both of them distribute their new
  blocks to other miners, and both blocks are accepted because they
  are not in conflict, the two blocks are added to the two separate
  chains, in parallel. If a single-strand chain had been used then
  only one of the blocks could have been successfully added. The
  increased efficiency lies in the fact that two blocks are added to
  the distributed ledger instead of just one.

  \section{Discussion}

  The multi-strand approach discussed in this paper allows increased
  efficiency because new valid blocks can be added in parallel, which
  becomes possible when using a set of semi-independent chains instead
  of a single chain.

  \subsection{Sealing blocks using signatures versus using a second proof-of-work phase}

  A previous version of this paper suggested to use a second second
  proof-of-work phase after the ticket creation, as a way to seal the
  block and protect from the risk of another node hijacking the ticket
  and using it to create a different block. A significant problem with
  that suggestion is that it is unclear how the difficulty for the
  second proof-of-work phase would be chosen. Miners can have widely
  varying CPU power, and a choice of difficulty that might be
  reasonable for one miner would be inadequate for others. Therefore,
  the method presented in Section~\ref{section:ticket_based_method}
  uses only one proof-of-work (for ticket creation) and employs a
  signature to allow other nodes to verify that the block was created
  by the ticket creator, thereby establishing the necessary connection
  between the proof-of-work and the corresponding block.

  \subsection{Related work}
  
  The idea of using several chains to increase efficiency is not new;
  one example of previous work employing this idea is the ``Parallel
  Chains'' work of Fitzi et al~\cite{ParallelChains2018}. However,
  although a similar idea is involved, their method differs
  significantly from the ticket-based approach outlined in the present
  paper. In~\cite{ParallelChains2018} the miner needs to prepare a
  metablock effectively containing candidate blocks for each of the
  $n$ chains, something that is not needed in our ticket-based
  approach which should therefore have better scalability for large
  values of $n$. The ticket-based method is also simpler in the sense
  that no special ``synchronization chain'' is involved, all $n$
  chains are treated in the same way, none of them is special and no
  other synchronization between chains is needed beyond the inclusion
  of last block hashes in the tickets.

  Another related work employing parallel chains is~\cite{OHIE2018}
  which however differs from our ticket-based method in
  that~\cite{OHIE2018} requires a miner to create the complete block
  without knowing which chain the new block will belong to. Our
  ticket-based method employs a two-phase approach where the ticket is
  created without knowing the chain index, but when the actual block
  is created the chain index is known, which is important as it allows
  the miner to select which transactions to include in the block
  knowing which chain the block will belong to.

  DAG-based protocols such as~\cite{GHOSTDAG2018} have some similarity
  with the present work in the sense that they allow multiple blocks
  to be generated in parallel. However, such methods are much more
  complicated than the simple scheme we outlined here because they
  involve complicated relationships between blocks where a block can
  have multiple predecessors leading to issues such as the
  ``late-predecessor'' phenomenon that can affect security and
  performance~\cite{SecurityPerformanceTradeoff2023}. In our
  ticket-based method, determining if a new block is valid is much
  simpler as it only requires access to the chain that block belongs
  to, nothing more. Note that although ticket creation involves hashes
  of blocks from all chains, validation of a ticket only requires
  checking one of those hashes, so there is in our case only one
  predecessor block.

  \subsection{Security}

  An attacker who wants to target one specific chain would want to
  create tickets specifically for the targeted chain, but that is not
  possible since the chain index only becomes known after a ticket is
  complete. So, the attacker is prevented from targeting a specific
  chain. A successful attack becomes possible only if the attacker
  controls more than half of the whole network, just as in the
  single-strand case.

  One drawback of the ticket-based multi-strand method as described in
  Section~\ref{section:ticket_based_method} is that a malicious miner
  who created a valid ticket can then attempt to cause confusion by
  creating multiple variants of the corresponding block, without
  redoing the proof-of-work. Since the miner has the private key for
  the ticket it is possible for the miner to sign several different
  blocks using the same ticket. Even though there is nothing to gain
  in the long run as only one of those blocks can survive, such
  behavior of malicious miners can increase the likelihood of
  conflicting versions of the most recently added block for a given
  chain. However, as soon as other blocks are added to the chain the
  block will be ``locked'' and protected by the proof-of-work in the
  added blocks. Just as in the single-strand case, for an attacker who
  controls less than half of the network, the probability of the
  attacker to catch up diminishes exponentially as more blocks are
  added.

  Another aspect worth considering is that since the multi-strand
  approach allows smaller block sizes to be used, it may be better
  suited for small miners which could be good for decentralization and
  security.

  \subsection{Implementation and usage}

  Implementation of the ticket-based method for a given single-strand
  blockchain (e.g. Bitcoin or Monero) should be relatively
  straightforward given that each of the $n$ chains can use the
  existing software with only slight modifications to include a ticket
  inside each block, and to add validation of the ticket inside a
  block when checking if a block is valid.

  Depending on the type of information stored, having separate chains
  may have drawbacks. In case of cryptocurrencies, having multiple
  chains means that each transaction can only involve outputs that
  belong to the same chain. So, it would be necessary for users to
  have a separate ``wallet'' for each chain. While this might seem
  awkward, it could potentially be handled by wallet software and the
  advantage of increased efficiency might be worth that extra
  complexity, especially if the goal is to handle the transaction
  rates needed to accommodate daily transactions of billions of users.

\end{document}